\documentstyle[preprint,aps]{revtex}
\newcommand{\be}{\begin{equation}}
\newcommand{\ee}{\end{equation}}
\newcommand{\bq}{\begin{eqnarray}}
\newcommand{\eq}{\end{eqnarray}}

\font\love=cmr10 scaled\magstep 1

\def\dsp{\displaystyle}

\def\cjsd{c_{{\bf j}, \sigma}^{\dagger}}
\def\cjs{c_{{\bf j}, \sigma}}

\def\cjupd{c_{{\bf j},\uparrow}^\dagger}
\def\cjdw{c_{{\bf j},\downarrow}}
\def\cjdwd{c_{{\bf j},\downarrow}^\dagger}
\def\cks{c_{{\bf k}, \sigma}}
\def\cksd{c_{{\bf k},\sigma}^\dagger}
\def\cjsp{c_{{\bf j},{\sigma} '}}
\def\ckspd{c_{{\bf k},{\sigma} '}^\dagger}
\def\nj{n_{\bf j}}
\def\nk{n_{\bf k}}
\def\nup{n_{{\bf j}, \uparrow}}
\def\ndw{n_{{\bf j}, \downarrow}}
\def\sj{\sum_{\bf j}}
\def\sjk{\sum_{<{\bf j},{\bf k}>}}
\def\diracij{\delta_{{\bf j}, {\bf k}}}
\def\sigsig{\delta_{\sigma , \, \sigma '}}
\def\ojk{O_{{\bf j}\,{\bf k}}}
\def\pjks{P_{{\bf j}\,{\bf k},\sigma}}
\def\qjks{Q_{{\bf j}\,{\bf k},\sigma}}
\def\rjks{R_{{\bf j}\,{\bf k},\sigma}}

\def\CcC{{\hbox{\love C\kern-.45em{\vrule height.66em width0.08em
depth-.04em
\hskip.45em }}}}
\def\RrR{{\hbox{\love I\kern-.17em{R}}}}
\def\ZzZ{{\hbox{\love Z\kern-.31em{Z}}}}
\def\NnN{{\hbox{\love {I\kern-.18em{N}}\kern-.18em{I}}}}
\def\IiI{{\hbox{\love I\kern-.19em{I}}}}

\begin{document}
\draft
\title{
Rigorous results on superconducting ground states for attractive extended
Hubbard models}
\author{Arianna Montorsi$^{*,\#}$, and David K. Campbell$^*$}
\address{
$^{*}$ Department of Physics, University of Illinois at Urbana-Champaign,
61801 USA \\
$^{\#}$ Dipartimento di Fisica and Unit\'a INFM, Politecnico di Torino,
I-10129 Torino, Italy}
\date{31 January 1995}
\maketitle
\begin{abstract}
We show that the exact ground state for a class of extended Hubbard models
including bond-charge, exchange, and pair-hopping terms, is the
Yang "$\eta$-paired" state for any non-vanishing value of the pair-hopping
amplitude, at least when the on-site Coulomb interaction is
attractive enough and the remaining physical parameters satisfy a
single constraint. The ground state is thus rigorously superconducting. Our
result holds on a bipartite lattice in any dimension, at any band filling, and
for arbitrary electron hopping.
\end{abstract}

\narrowtext

Interest in itinerant, strongly interacting electron systems has exploded
since the discovery of high-$T_c$ superconductors, for which the interplay
between itinerant magnetism and insulating behavior is believed to play a
crucial role. The Hubbard model\cite{HUB} provides the simplest description
of such systems, by assuming that the itinerant electrons interact only via
an on-site term, and it has therefore been intensively studied. The original
Bethe Ansatz solution of the one-dimensional model at half-fllling
\cite{LIWU} has in recent years been supplemented by a number of additional
rigorous results (see \cite{Lieb93} for a current review). For instance, for
the {\it attractive} ({\it i.e.}, ``negative U'') Hubbard model, long
studied as a model believed to have a superconducting ground state
\cite{MIRA}, recent articles have established the existence of
``off-diagonal long-range order'' (ODLRO \cite{YAN2}),  but these results
have been restricted either to $U\rightarrow -\infty$ \cite{SiSc}
or to bipartite lattices in which the number of sites on one sublattice is
not equal to that on the other \cite{SQ,Tian,Lieb93}. Indeed, although it is
known \cite{YAN1} that the Hubbard Hamiltonian has certain eigenstates (the
``$\eta$-paired'' states) that have non-zero pairing and ODLRO
and hence are superconducting, Yang has shown that these
states can never be the ground state for the pure Hubbard model on a
standard bipartite lattice \cite{YAN1}.

Apart from the Hubbard model itself, ``extended'' Hubbard models have
also attracted considerable recent interest, in part because it has been
recognized \cite{HIR,CAGA} that the additional interaction terms
(discussed but eventually neglected in the  original papers \cite{HUB})
could be relevant  in  stabilizing some novel ({\it e.g.}, ferromagnetic or
superconducting) phases.  For instance, Strack and Vollhardt  \cite{STVO}
proved rigorously  that the ground state for a class of extended  Hubbard
models is a saturated ferromagnet at half-filling for {\it any}
non-vanishing value of the exchange term, provided that the Coulomb
repulsion  is strong enough.

In this Letter we show that in appropriate regions of the parameter space
of a class of extended Hubbard models, the superconducting phase is the
stable ground state for {\it any} non-vanishing value of the pair-hopping
term  \cite{HIR,CHSU}. More  precisely, by simultaneous use of Yang's
states and of a suitable generalization of Strack and  Vollhardt's
techniques, we prove rigorously that the ground-state of the  extended
Hubbard model at any band filling and for average magnetization $m=0$  is
superconducting,  provided that the on-site Coulomb interaction is
attractive enough and that  the remaining physical parameters satisfy a
single reasonable constraint. Moreover, by a particle-hole transformation,
we map our result into  a sufficient criterion for stability of
ferromagnetism in the repulsive  case at  half-filling and for arbitrary
$m$,  which criterion is more general than  the one given in \cite{STVO}.
Indeed, this relation between  superconductivity for negative $U$ and
magnetism for positive $U$ can  readily be understood in terms of the
$SO(4)$ symmetry  \cite{SO4} underlying the model. Importantly, in contrast
to other recent work \cite{DEKO},  the supersymmetric condition $t=X$ (see
(\ref{ham}) below) is in general neither required  nor fulfilled  by our
superconducting ground state.

Our extended Hubbard Hamiltonian reads \cite{HUB}:
\widetext
\bq
H &=& - t\sjk\sum_\sigma \cjsd \cks +  U \sj \nup\ndw  + X \sjk\sum_\sigma
(n_{{\bf j},-\sigma} + n_{{\bf k},-\sigma})  \cjsd \cks \nonumber \\ &+&
{V\over 2}\sjk  \nj \nk + {W\over 2} \sjk \sum_{\sigma,\sigma'} \cjsd \ckspd
\cjsp \cks + {Y\over 2} \sjk \sum_\sigma \cjsd c_{{\bf j},-\sigma}^\dagger
c_{{\bf k},-\sigma} \cks \quad , \label{ham}
\eq
\narrowtext
where $\cjsd , \cjs \,$ are fermionic creation and annihilation operators ($
\{ \cjsp, \cks \} = 0 \,$, $\, \{ \cjs , \ckspd \} = \diracij\, \sigsig
{\IiI}$, $\, n_{{\bf j},\sigma} \doteq \cjsd \cjs$, $\nj = \sum_\sigma
n_{{\bf j},\sigma}$ ) on a $d$-dimensional
lattice $\Lambda$ (${\bf j} , \, {\bf k} \in \Lambda$, $\sigma \in \{
\Uparrow , \Downarrow \}$), and $< {\bf j} , \, {\bf k} >$ stands for
nearest neighbors ({\it n.n.}) in $\Lambda$. In (\ref{ham})
the first term represents the band energy of the electrons, and the
remaining  terms describe their Coulomb interaction energy in a narrow
band  approximation \cite{HUB}: $U$ parametrizes the on-site diagonal
interaction, $V$ the neighboring site charge interaction, $X$ the
bond-charge interaction, $W$ the exchange term, and $Y$ the pair-hopping
term. An explicit evaluation of the relative size of these contributions
-- all generated from onsite and {\it n.n.} matrix elements of the Coulomb
interaction -- was already given in \cite{HUB}.  It is
worth emphasizing that most of the following analysis can be extended in a
straightforward way to the case in which the  interactions in (\ref{ham})
are not confined to neighboring sites. However, to avoid
cumbersome notation, we have chosen to limit ourselves to the present case.

The exchange and pair-hopping terms in (\ref{ham}) can be written more
conveniently
in terms of the conventional spin and pseudo-spin operators $S_{\bf
j}^{(\alpha)}$ and $\tilde S_{\bf j}^{(\alpha)}$, $\alpha = x,y,z$,
\bq
S_{\bf j}^{(+)} &=& \cjupd\cjdw \; , \; S_{\bf j}^{(-)} =  (S_{\bf
j}^{(+)})^\dagger \; , \; S_{\bf j}^{(z)} = {1\over 2} (\nup -\ndw)
\nonumber \\
\tilde S_{\bf j}^{(+)} &=& (-)^{|{\bf j}|} \cjupd\cjdwd \; , \; \tilde S_{\bf
j}^{(-)} =  (S_{\bf j}^{(+)})^\dagger \; , \; \tilde S_{\bf j}^{(z)} =
{1\over 2}  (\nup + \ndw-1) \label{su2}\\ S_{\bf j}^{(\pm)} &\doteq& S_{\bf
j}^{(x)}\pm  i S_{\bf j}^{(y)} \quad , \quad \tilde S_{\bf j}^{(\pm)} \doteq
\tilde S_{\bf j}^{(x)}\pm i \tilde S_{\bf j}^{(y)} \nonumber \quad.
\eq
which are known to generate two orthogonal $su(2)$ algebras.
For a {\it bipartite} lattice with an even number of sites, the second line of
(\ref{ham}) reads then
\be
2 V'\sjk \tilde S_{\bf j}^{(z)} \tilde S_{\bf k}^{(z)} -
W \sjk {\bf S}_{\bf j} \cdot {\bf S}_{\bf k} - Y \sjk \tilde {\bf S}_{\bf j}
\cdot \tilde {\bf S}_{\bf k} + C
\quad , \label{hams}
\ee
where $V'=V-\dsp{1\over 2} (W-Y)$ and $C = \dsp{1\over 2} ( V - {W\over 2}) q N
(2 n -1)$, with $q$ number of nearest neighbors in $\Lambda$, $N$ number of
sites, and $n$ average electron number per site.

$H$ can be easily seen to commute with $su(2)_f$,
generated by $S_\alpha = \dsp{\sj  S_{\bf j}^\alpha}$, $\alpha=x,y,z$. This
observation is in fact  relevant in recognizing the $(N + 1)$-fold degenerate
saturated ferromagnetic  state $|\psi_m>$,
\be
|\psi_m> \doteq (S_+)^{(n+m)N\over 2} |\downarrow \downarrow \dots \downarrow>
\quad, |m|\leq \min\{1,n\} \quad , \label{ferro}
\ee
as eigenstate of $H$ at $n=1$ (half-filling) and average magnetization per
site $m$. Analogously, one can construct
$su(2)_{sc}$, generated  by $\tilde S_\alpha = \dsp{\sj \tilde S_{\bf
j}^\alpha}$. The latter does  not commute with $H$. Nevertheless, it is
easily checked that the states  -- known as $\eta$-pairs \cite{YAN1} --
defined by
\be
|\eta_n>\doteq  (\tilde S_+)^{nN\over 2} |0> \quad ,\quad 0\leq n\leq 2
\quad ,
\label{eta}
\ee
with $|0>$ being the electron vacuum, are eigenstates of $H$ at band filling
$n$  with $m=0$, provided that $V'=0$. Interestingly enough, these states
are  precisely those which Yang has proved to exhibit ODLRO, which  property
has been shown to imply both Meissner effect and  flux quantization
\cite{NISU}, {\it i.e.} superconductivity. Notice that the $\eta$-pairs
(\ref{eta}) differ from the pairs defined in  \cite{DEKO} by a factor of
$(-)^{|{\bf j}|}$; this will be essential in the following.

Assuming henceforth $V'=0$, we want to investigate under which circumstances
$|\eta_n>$ is  indeed the ground state. We proceed by rewriting, by means of
suitable operator identities, the hopping and bond-charge repulsion  terms
as sums of positive definite operators  having zero eigenvalue on
(\ref{eta}) plus contributions which simply renormalize the other terms in
the Hamiltonian, which now reads
\widetext
\bq
H &=& - U'\sj [S_{\bf j}^{(z)}]^2 - W' \sjk {\bf S_j} \cdot {\bf S_k} -
Y' \sjk \left( \tilde {\bf S}_{\bf j}\cdot \tilde {\bf S}_{\bf k} -
\frac{1}{4} \right )  + J \sjk S_{\bf j}^{(z)} S_{\bf k}^{(z)} + C'
\label{ham1}\\
&+&\sjk\left\{ |t-X| |\gamma| \ojk^\dagger \ojk  + \sum_\sigma\left [
|t-X||1-\gamma|  \pjks^\dagger\pjks + |X| (\qjks\qjks^\dagger +
\rjks^\dagger\rjks )\right] \right\} \, , \nonumber
\eq
\narrowtext
where $ U' = 2 \{ U + q[|t-X| (|\gamma |\alpha^2 + |1-\gamma| \beta^2) +4
|X|]\}$, $W' = W - \beta^2 |t-X||1-\gamma|$,
$Y'=Y-|t-X|\dsp{[{2|\gamma|\over{\alpha^2}} +{|1-\gamma|\over{\beta^2}}]} $,
$J=|t-X|[\dsp{2 |\gamma| (\alpha^2 + {1\over{\alpha^2}}) - |1-\gamma| (\beta^2
+ {1\over{\beta^2}})}]$, and $C'= C-\dsp{{1\over 2} ( U n+Y {q\over 2}) N}$.
Here $\alpha\neq 0$, $\beta\neq 0$, and $\gamma$ are free paramaters, and
\bq
\ojk&=& \alpha(S_{\bf j}^{(z)} - S_{\bf k}^{(z)}) +
{\epsilon\over\alpha} \left (c_{{\bf k},\uparrow}^\dagger  c_{{\bf
j},\uparrow} + c_{{\bf j},\downarrow}^\dagger c_{{\bf  k},\downarrow} \right
)\quad,\nonumber\\
\pjks&=&{1\over 2}\left [ \beta (S_{\bf j}^{(\sigma)} - S_{\bf k}^{(\sigma)}) +
{\eta\over\beta}\left ( \cksd c_{{\bf j},-\sigma} - \cjsd
c_{{\bf k},-\sigma} \right)\right ]\quad, \nonumber\\
\qjks&=&{1\over\sqrt{2}}\left( n_{{\bf j},-\sigma}\cjs +\theta
n_{{\bf k},-\sigma}\cks\right) \quad, \nonumber\\
\rjks&=& {1\over\sqrt{2}}\left[(1-n_{{\bf j},-\sigma})\cjs +\theta (1-
n_{{\bf k},-\sigma})\cks \right] \quad,   \label{op1}
\eq
with $\epsilon=sgn[(t-X)\gamma]$, $\eta=sgn[(t-X)(1-\gamma)]$, and
$\theta=sgn[X]$.

For $U'\leq 0$ and $Y'\geq 0$ the lower bound of the on-site and pair-hopping
terms in (\ref{ham1}) is zero,  which coincides with their eigenvalue on the
states $|\eta_n>$. Moreover,
\be
\ojk|\eta_n>=\pjks|\eta_n>=\qjks^\dagger|\eta_n>=\rjks|\eta_n>=0  \quad ,
\ee
and -- $H$ being a positive definite form in these operators --
$|\eta_n>$ is  the ground state also for the second line of (\ref{ham1}).
The freedom in the choice of $\alpha$, $\beta$, and
$\gamma$ permits two of them to be fixed so that $W'=0=J$. In this case, one
finds that the state $|\eta_n>$ is the ground state with energy
$E_{\it gs} = \dsp{1\over 2} (U-q Y) n N$ whenever $U'\leq 0$, and $Y'\geq 0$.
In fact, we can obtain an even larger region of values of $U$ for which
$|\eta_n>$ is the ground state if, instead of fixing
$\alpha$, $\beta$, and $\gamma$ as above, we first express both the
Ising-like  term and the ferromagnetic exchange term in (\ref{ham1}) through
the following  operator identities,
\bq
-W' \sjk {\bf S_j} \cdot {\bf S_k} &=& |W'|\left (\sjk B_{\bf j \,k}^\dagger
B_{\bf j\, k} - 2 q \sj [S_{\bf j}^{(z)}]^2 \right ) -W'\sjk S_{\bf j}^{(z)}
S_{\bf k}^{(z)}\nonumber\\
J'\sjk S_{\bf j}^{(z)} S_{\bf k}^{(z)} &=& |J'|\left ({1\over  2} \sjk
[S_{\bf j}^{(z)}+
sgn(J')S_{\bf k}^{(z)}]^2 - q \sj [S_{\bf j}^{(z)}]^2\right) \quad , \label{id}
\eq
with
\be
B_{\bf j\, k} = {1\over\sqrt{2}}\left (c_{{\bf j},\uparrow}^\dagger c_{{\bf j},
\downarrow} - sgn(W')c_{{\bf k},\uparrow}^\dagger c_{{\bf k},\downarrow}
\right ) \quad , \label{op2}
\ee
and $J'=J-W'$. The identities (\ref{id}) -- when inserted in (\ref{ham1}) --
further renormalize the coefficient of the Coulomb interaction term ({\it i.e.}
$\sj [S_{\bf j}^{(z)}]^2$), so that the inequalities which have to be
satisfied in order that $|\eta_n>$ be the ground state now read
\be
U'+ q (2 |W'|+|J'|) \leq 0\quad ,\quad Y'\geq 0 \quad , \label{ineq}
\ee
and are still functions of $\alpha$, $\beta$, and $\gamma$. Eliminating
$\alpha$ using the second of Eqns. (\ref{ineq}), one is left with a single
inequality for $u\doteq\dsp{U\over q t}$,
%\bq
%- {U\over q} \geq &&4 X +|t-X|\left [ \frac{2\gamma^2
%|t-X|}{Y\beta^2-|1-\gamma| |t-X|}\right] + \left|W-\beta^2|t-X|
%|1-\gamma|\right| \nonumber\\
%&&\left|{1\over 2}(Y-W)+\frac{2 (t-X)^2\gamma^2 \beta^2}
%{Y\beta^2-|1-\gamma| |t-X|}-\frac{|1-\gamma| |t-X|}{\beta^2}\right|
%\quad ,\label{ueq}
%\eq
\be
u\leq
-\left\{4x+|1-x|\left[\frac{2\gamma^2}{B}+\frac{(1-\gamma)^2}{y-B}+
\left|\frac{w}{|1-x|}-\frac{(1-\gamma)^2}{y-B}\right|+\left
|\frac{2\gamma^2}{B}+B-\frac{w+y}{2|1-x|}\right|\right]\right\},
\label{ueq}
\ee
where $x$, $y$, and $w$ are $X$, $Y$, and $W$ expressed in units of $t$, and
$B=\dsp\frac{y}{|1-x|} -\dsp{|1-\gamma|\over\beta^2}$, $0<B<y$. Equation
(\ref{ueq}) can be optimized by fixing the remaining  parameters  $\beta$
(through $B$) and $\gamma$ variationally. The result gives a {\it
sufficient} condition for the superconducting state $|\eta_n>$ to be  stable
ground state. This ground state is {\it unique} for $Y'>0$, for in  that
case it is the unique ground state of the pair-hopping term in
(\ref{ham1}). The constraint $V'=0$ ({\it i.e.}  $V=\dsp{1\over 2}(W-Y)$)
must be satisfied. The above rigorous result holds for any bipartite
lattice, in  any dimension, and importantly for {\it arbitrary} values of $x$.
In particular, for $v=x=y=w=0$ one finds directly from (\ref{ueq})
that the pure
Hubbard model has a superconducting ground state with ODLRO at least for
$u=-\infty$, in agreement with \cite{SiSc}.

The final explicit form of the general result (\ref{ueq}) is too long  to be
written in the present letter, and will be reported elsewhere  \cite{MOPI}.
Here we simply plot in Fig. \ref{Fig1} the  actual boundary  of our rigorous
superconducting region in the $u$ $vs$ $y$ plane at different $x$ values.
The inequality  (\ref{ueq}) and Fig. \ref{Fig1} show that {\it any}
$y\neq 0$ there is a region  of $u$ values for which the system is
superconducting, and its size  increases with increasing $y$, at least for
$y\leq|1-x|$.  It is quite natural that a  non-vanishing value of $y$ can
stabilize the superconducting phase, in that  it removes the degeneracy of
the expectation value of the Hamiltonian on  states $|\Phi_n(\phi)> = [\sj
e^{i {\bf \phi} \cdot {\bf j}}  (c_{{\bf j},\uparrow}^\dagger c_{{\bf
j},\downarrow}^\dagger)]^{n {N\over  2}}$, which, for arbitrary $\phi\neq
\pi$ and $x\neq 1$, are not eigenstates  of $H$. As pointed out by Yang in
\cite{YAN1}, this observation implies that  $|\eta_n>\equiv|\Phi_n(\pi)>$
can not possibly be the ground state at $y=0$.  On the contrary, for $x=1$
and $y=0$,  $|\Phi_n(\phi)>$ is an eigenstate of $H$ for all $\phi$  and can
in principle be the (degenerate) ground state. Indeed, from  (\ref{ueq}) we
see that this is the case at least for $u\leq - 4$.  For $y\neq 0$, $x\neq
1$, $<\Phi_n(\phi)|H|\Phi_n(\phi)>$ becomes a function of  $\phi$, in fact
minimized by $\phi=\pi$, and $|\Phi_n(\pi)>$ turns out to  be the ground
state at least in the region of $U$ values satisfying  (\ref{ueq}). Finally,
for $x=1$ and $y\neq 0$, there are two choices of  $\phi$ which correspond
to eigenstates of $H$, $\phi=\pi$ (for  $V=\dsp{1\over 2}(W-Y)$) and
$\phi=0$ (for $V=\dsp{1\over 2}(W+Y)<0$). The  first corresponds to the
ground state at least for $u\leq - (4 + y + w)$  (see (\ref{ueq})), whereas
the region of stability of the solution corresponding to the  second was
already discussed in \cite{DEKO}.

The relation (\ref{ueq}) has a solution only for negative values of $u$.  On
the other hand, the physics of high-$T_c$ materials suggests that the actual
value of  the on-site electron interaction is strongly repulsive. Even if
the  electron-phonon coupling reduces the effective value of the
Hubbard interaction \cite{ALRA}, its sign is still expected to be positive.
However, one should keep in mind two points. First, (\ref{ueq}) is a {\it
sufficient}  condition, and thus does not eliminate the possibility of
having $|\eta_n>$ as  ground state even when it is not fulfilled. Again, the
easier case $x=1$  helps clarify this point. There, an exact solution in 1-d
\cite{E1D} shows that ODLRO and superconductivity still survive as part of
the degenerate ground state up to moderately positive values of $u$, which
values are now band-filling dependent. We thus expect this behavior
to persist even when the condition $x=1$ is relaxed. In particular,
it would be extremely interesting to work out an exact Bethe Ansatz solution
of  (\ref{ham}) in $d=1$, at least at $x=0$ (for $u=w=0$ the latter would be
the  superconducting "$t-Y$" model, which is the particle-hole
transformation  of the {\it ferromagnetic} $t-J$ model \cite{PULU}). Second,
and very importantly, knowing  rigorously both the superconducting
nature of the ground state and
its  explicit form in any dimension provides powerful benchmarks for the
approximate methods required to examine more realistic models.

Apart from the superconducting solution, our expression (\ref{ham1}) for $H$
allows us also to recognize a region of the parameter space characterized by
ferromagnetic order. Indeed, it is easily seen that the
state $|\psi_m>$ of (\ref{ferro}) can be obtained from the state
$|\eta_n>$ of (\ref{eta}) by the following unitary particle-hole
transformation:
\be
c_{{\bf j}, \uparrow}\rightarrow c_{{\bf j},\uparrow} \quad , \quad
c_{{\bf j}, \downarrow}^\dagger\rightarrow (-)^{|{\bf j}|}c_{{\bf
j},\downarrow}
\quad . \label{parthole}
\ee
The same transformation maps $S_{\bf j}^{(\alpha)}$ into $\tilde  S_{\bf
j}^{(\alpha)}$, and the consequences for $H$ as given by (\ref{ham1}) can
be worked out directly. Let us call the  transformed Hamiltonian $\tilde H$.
For $X=0$, $\tilde H$ is still an extended Hubbard model, in which the
on-site Coulomb  repulsion  term has opposite sign, $W$ and $Y$ have
exchanged their roles (the first  becoming the pair-hopping amplitude, and
the second the exchange coupling), and the operators of
(\ref{op1})-(\ref{op2}) have been redefined accordingly.  Moreover, an
arbitrary neighboring-site Coulomb repulsion  term can be added to $\tilde
H$, as now it simply renormalizes the  coefficient $J$ of the Ising-like
term in (\ref{ham1}). The discussion following (\ref{ham1}) can be used to
examine the conditions under which the saturated ferromagnetic state
$|\psi_m>$ is the ground state of $\tilde H$. A  straightforward calculation
shows that the result is identical (in form) to the one given in
(\ref{ueq}), apart from the sign of $u$, the inequality hence becoming a
lower bound for positive $u$.  Further, now it is the
exchange coupling which  cannot be zero in order to have a stable
ferromagnetic phase.  This result is in full agreement with \cite{STVO}. In
fact, our lower bound can easily be seen to coincide with expression (6) of
\cite{STVO} for $\gamma=0$, whereas it is lower than that if $\gamma$ is
fixed variationally. Again, a more complete discussion of this case will be
given  elsewhere \cite{MOPI}. Notice that in \cite{STVO}  the freedom  in
the polarization of the saturated ferromagnet was not  explicitly
incorporated, which hid the power of the particle-hole  transformation.

In summary, we have shown rigorously that a large class of extended
Hubbard models on bipartite lattices has a superconducting ground state for
negative $U$ and non-vanishing pair-hopping amplitude. The conditions
derived here are sufficient, depend in a trivial way on the dimension, and
do not depend at all on the band filling. Of course, this does not exclude that
a superconducting ground state can exist even for moderate positive
values of $U$. If this is the case, we expect that the dimension and the
band filling should become crucial, as happens for instance in the $t=X$ case.
Work is in progress along these lines.

The authors gratefully acknowledge the hospitality of the Aspen Center for
Physics, where the present work was conceived. We also thank Fabian Essler,
Holger Frahm, Vladimir Korepin, Elliott Lieb, Andreas Schadschneider, and
Dieter Vollhardt for interesting discussions. This work was supported by the
NATO  fellowship program (AM) and the University of Illinois.

\begin{figure}
\caption{The boundary of the superconducting phase in the $-u$-$y$ plane
for different $x$ values,
$x=0,\,.5,\, 1$ (represented by continuous, dashed, and dot-dashed lines,
respectively), for $v=0$, $w=y$.}
\label{Fig1}
\end{figure}
\end{document}